\documentstyle[prl,aps,epsf]{revtex}

\begin{document}
\title{Symmetries, invariants and cascades in a shell model of turbulence
}
\author{P. D. Ditlevsen\\
The Niels Bohr Institute, Department for Geophysics,\\
University of Copenhagen, Juliane Maries Vej 30,\\
 DK-2100 Copenhagen O, Denmark.}
\date{\today}
\maketitle
\begin{abstract}
Reduced wavenumber models of turbulence, shell models, show
cascade processes and 
anomalous scaling of correlators which might be analogous to 
what is observed in Navier-Stokes (N-S) turbulence. The scaling
properties of the shell models depends on the specific
symmetries and invariants of the models. 
A new shell model is investigated. It is argued that 
this model might have a closer 
resemblance than the standard GOY model 
to the N-S turbulence.
The
new shell model coincides
with the Sabra model 
proposed by L'vov et al. \cite{lvov}, for a
specific choice of the free parameters of their model. 
The model has for this choice of parameters, besides the
energy and the 'helicity', a cubic inviscid invariant.
\end{abstract}
\pacs{PACS appear here.}

\section{Introduction}
The connection between Kolmogorovs 1941 (K41) scaling theory of
turbulence and the N-S equation is through 
the four-fifth law \cite{Frisch}.
The four-fifth law is one of the few rigorous result regarding 3D hydrodynamic turbulence,
connecting the second and third order longitudinal velocity structure
functions. The equation is not closed, but in the inertial range, $\eta
\ll r \ll L$, $\eta$ being the inner (Kolmogorov) scale and $L$ the
outer (integral) scale, we have $S_3(r) = - (4/5) \epsilon r$, where
$\epsilon$ is the mean energy dissipation. 
This implies  
$\zeta(3)=1$, where $\zeta(p)$ is defined from the scaling of the
longitudinal structure functions, 
$S_p(r) = \langle \delta v_{\parallel}(r)^p\rangle \sim r^{\zeta(p)}$. 
The scaling of all other structure functions are observed
to deviate from the K41 prediction, which from dimensional counting would imply
$\zeta(p)= p/3$. The deviation from the K41 prediction is called anomalous
scaling exponents, referring to the intermittent nature of the energy dissipation,
where the energy is inhomogeneously dissipated. 

As was noted recently \cite{Chkhetiani} the
inviscid conservation of helicity leads to another exact scaling law
for a third order correlator associated with the flux of helicity, provided
an inertial range for helicity flux exists \cite{D&G}. 
Similarly for any inviscid invariant an exact scaling law can be derived
for the correlators associated with the spectral flux of such a quantity
provided there is an inertial range separating sources and sinks for this
quantity. This is the case for the new shell model, where the existence
of a third order inviscid invariant leads to a scaling law for a 
fourth order correlator. This has been argued to be a calculation of an
anomalous scaling exponent \cite{lvov1}. This is, however, not an anomalous
scaling exponent in the afore mentioned sense, it would still be present
in the case of an (imagined) homogeneous non-intermittent flow. On the contrary, the
scaling of this specific fourth order correlator will be 'normal' in the
same sense as $\zeta(3)=1$i from the four-fifth law. 

The resent interest in shell models is mainly that
they show numerically the same type of 
intermittent 
behavior as seen in 3D turbulence \cite{jpv}. 
The simplicity of the shell models makes it possible to
calculate the anomalous scaling exponents with high accuracy,
which is still an undo-able task for the N-S equation.
so in this sense shell models might
prove useful as the starting point for exact results regarding
calculating scaling exponents. The introduction of the
Sabra model \cite{lvov} as superior to the GOY model \cite{Gledzer} 
was motivated by improvements in accuracy with respects to 
numerical determination of scaling exponents. Here the emphasis will
be on the similarity with the N-S equation and the cascade properties
of the third order invariant. In fact the model will not exhibit
a cascade of the third order quantity, which is easily seen from a scaling
argument and confirmed in a numerical simulation. Thus it will be shown that
the scaling exponent obtained in \cite{lvov1} is irrelevant.
 
\section{The GOY model}

The GOY model has built in the K41 scaling in the sense that
the K41 scaling is a fixed point of the model. Furthermore, it
has an unfortunate modulus(3) symmetry in shell numbers which
has no resemblance in the N-S equation and which makes 
a precise numerical determination of scaling exponents difficult
\cite{Kadanoff,lvov1}. The modulus(3) symmetry, furthermore , introduces
artificial long range (in $k$-space) correlations with no analogs in
the N-S equation. The GOY model has 2 inviscid invariants,
the energy and a second non-positive definite quantity dimensionally
equivalent to the helicity in 3D N-S turbulence.
This 'helicity' only vaguely resembles the helicity in the 
N-S fluid, and it has been argued that it leads to an anomalous
scaling behavior of the GOY model different from the
mechanisms for intermittency in N-S turbulence \cite{Gat}.
A review of the main differences between the GOY model and the Sabra
model is given in Ref. \cite{lvov}. 

\section{The new model}
The shell model, defined in the following, can be
regarded as a special case of the Sabra model
introduced by L'Vov et al. \cite{lvov}.
It has the same two quadratic
inviscid invariants, energy and 'helicity' as the GOY model. Furthermore it has
one cubic inviscid invariant.
The energy is the only positive
invariant. 
Like for the N-S equation, and in contrast to the GOY model, 
the K41 scaling is not a fixed
point of the model. 

The model is in the usual way defined by a set of
exponentially spaced one-dimensional wavenumbers $k_n=k_0 \lambda^n$
for which we have $u_n$ as the complex shell velocity for shell $n, (n=1,N)$,
The form of the governing equation for the new model is motivated from two demands.
Firstly, the momenta involved in the 
triad interactions must add up to zero like in the N-S equation.
Secondly, the complex conjugations involved in the non-linear term
should be the same as for the terms in the spectral N-S equation involving
triades for which the moduli of the wave vectors
fall within three consecutive shells.
This together with the usual construction of local interactions in $k$-space, inviscid
conservation of energy and fulfillment of Liouvilles theorem gives
the equation of motion for the shell velocities,

\begin{eqnarray}
(d/dt + \nu k_n^2)u_n=\nonumber \\
i(k_nu_{n+1}^*u_{n+2}-\epsilon k_{n-1}u_{n-1}^*u_{n+1}+(1-\epsilon)k_{n-2}
u_{n-1}u_{n-2})+f_n
\label{du}
\end{eqnarray}
where
$\nu$ is the viscosity and $f_n$ is the external forcing. The
forcing would as in the GOY model typically be taken to be active only for
some small wave numbers, ex. $f_n=f \delta_{n,4}$. 

Boundary conditions can be specified in the usual way 
by the assignment $u_{-1}=u_0=u_{N+1}=u_{N+2}=0$.

The first requirement is fulfilled if the wave numbers $k_n$ are defined as a Fibonacci sequence, 
$k_{n}=k_{n-1}+k_{n-2}$. 
The choice of a
Fibonacci sequence for the momenta leads to a model with the
shell spacing uniquely being the golden ratio $g$,  since for any choice
of $k_1, k_2 (k_1\le k_2)$ we have $k_n/k_{n-1}\rightarrow g$ for
$n\rightarrow \infty$. 
So this corresponds to the
usual definitions of the shell wave numbers with the golden 
ratio as shell spacing 
for $k_1=1, k_2 =(\sqrt{5}+1)/2 \equiv g$ \cite{golden}.
The golden ratio $g=(1+\sqrt{5})/2$ plays a key role
in the symmetries of shell models. 


With this formulation the shell spacing
is not a free parameter of the shell model. However, using the
definition by L'vov et al of $k_n=g^n$ being a 'quasi-momentum' we
shall keep the shell spacing, $\lambda$, as a free parameter,
$k_n = \lambda^n$.

If we interpret the momenta, $k_n$, as representative of 
the modulus of the wavevector, ${\bf k}_n$, in 2D or 3D, the
triangle inequality implies $k_n+k_{n+1}\ge k_{n+2}$ so the
Fibonacci sequence corresponds in this sense 
to moduli of three parallel wavevectors.
Note that for a shell spacing $\lambda > g$ (as the usual
choice $\lambda =2$) the triangle inequality is violated.
This means that we cannot interpret the usual shell model 
interactions as representative interactions between waves
within three consecutive shells, since no such triplets
of wavenumbers constitutes triangles.

In order to give meaning to the notion of closing the triades, we
define negative momenta, $k_{-n}\equiv -k_n$, and assign the 
velocity, $u_{-n}=u_n^*$,
to these momenta. 
(The model still only has $2N$ degrees of freedom,
represented by the $N$ complex velocities). 
Note that (\ref{du}) is also fulfilled for the negative momenta,
which is why the prefactor must be $'i'$.
With this notation we can rewrite (\ref{du}) as

\begin{equation}
(d/dt + \nu k_n^2+)u_n=i k_n \sum_{k_l<k_m}\tilde{I}(l,m;n)u_lu_m +f_n,
\label{du1}
\end{equation}
where the sum is over positive and negative momenta,
and all the dimensionless interaction coefficients have the simple form,
$\tilde{I}(l,m;n)= I(l,m;n) \delta_{k_n+k_l+k_m,0}$ with
$I(l,m;n)=\delta_{n-2,l}\delta_{n-1,m}-
(\epsilon/\lambda)\delta_{n-1,l}\delta_{n+1,m}+
((1-\epsilon)/\lambda)\delta_{n+1,l}\delta_{n+2,m}$.
From this formulation it is clear why the complex conjugations
in this model are exactly as in (\ref{du}), they arise from the
closing of the triades in the same way as for the spectral N-S equation.
As noted in \cite{lvov} this is the main difference 
from the GOY model.

\section{Inviscid invariants}
It can easily be shown that there are only second order (quadratic)
invariants of the form, $\sum \xi^nu_nu_{-n}$.
The inviscid conservation of these quadratic
invariants is obtained from,

\begin{eqnarray}
\frac{d}{dt}\frac{1}{2}\sum_{|n|\le N} \xi^n u_nu_{-n} = \\
i \sum_{k_l+k_m=k_n}\xi^n k_nI(l,m;n)u_lu_mu_{-n} =
\nonumber \\
i\sum_{3\le |n|\le N} \xi^{n-2}k_{n-2}(1-\epsilon\xi-(1-\epsilon)\xi^2)
u_{n-2}u_{n-1}u_{-n} \nonumber \\
=0. \nonumber
\end{eqnarray}
so 
exactly as for the GOY model we obtain the equation,

\begin{equation}
1-\epsilon\xi-(1-\epsilon)\xi^2=0
\label{e2}
\end{equation}
with
the two
solutions $\xi=1$ and $\xi = 1/(\epsilon-1)$.
The first corresponds to energy conservation, with

\begin{equation}
E=\sum E_n =\frac{1}{2}\sum u_nu_{-n},
\end{equation}
and the second to 'helicity' conservation (for $\epsilon < 1$), with

\begin{equation}
H=\sum H_n =\frac{1}{2}\sum_{|n|\le N} \big(\frac{1}{(\epsilon-1)}\big)^{n} u_nu_{-n}.
\end{equation}

With the definition of negative momenta the
only possible $p$th order invariants with terms
$u_{i_1}...u_{i_p}$ must have the associated momenta 
summing to zero. Thus for any invariant,

\begin{equation}
\sum_i\xi_i^1  u_{j_1}...u_{j_p}+ ... + \xi_i^l u_{l_1}...u_{l_p} + c.c.,
\label{psum}
\end{equation}
the corresponding momentum vectors must add up to zero, 
$k_{j_1}+ ... + k_{j_p} = k_{l_1}+ ... + k_{l_p} =0$.
This can easily be seen by differentiating (\ref{psum}) with respect to
time using (\ref{du1}).
The only possible third order term is

\begin{equation}
G=\sum_n G_n =\sum_n\xi^n ( u_{n-1}u_{n}u_{-(n+1)} + c.c.)
\label{E3}
\end{equation}
By taking the derivative with respect to time using (\ref{du1}) we
obtain a set of equations similar to (\ref{e2}):

\begin{eqnarray}
\epsilon \xi + \lambda =0 \nonumber \\
(1-\epsilon)\xi^2-\lambda^2=0 \nonumber \\
(1-\epsilon)\xi+\epsilon\lambda =0
\end{eqnarray}
These equations are fulfilled if $1-\epsilon-\epsilon^2=0$ 
with $\xi=-\lambda/\epsilon$.
The two solutions are, $\epsilon = -g $ and $\epsilon = 1/g$.

From a similar analysis it can also be seen that there are
no invariants with $p>3$.

\section{Parameter space}
With $\epsilon$ and $\lambda > 1$ being the two free parameters
of the model the parameter space is represented in figure 1.
The hyperbola is the curve $\lambda=1/|\epsilon-1|$ corresponding
to the dimensionally correct helicity, $H=\sum (-1)^n k_n |u_n|^2$.
The dashed vertical lines corresponds to values where the third
order quantity $G$ is an inviscid invariant. The point, 
$(\epsilon,\lambda)=(1/g,g^2)$, marked with a square,
has both the usual helicity and $G = \sum (-1)^ng^{3n} u_{n-1}u_nu_{-(n+1)}$ 
conserved.
The point $(\epsilon,\lambda)=(2-g,g)$, marked with a diamond, does not have $G$ conserved 
 while the point $(\epsilon,\lambda)=(1/g,g)$, marked with a cross,  has $G$ conserved with $\xi=-g^2$ and
a helicity of the form, $H=\sum (-1)^n k_n^{1/2} |u_n|^2$.
The point $(\epsilon,\lambda)=(1/2,2)$, marked with a solid ball, is the point 
investigated by L'vov et al. corresponding to the values originally chosen for the
GOY model. In the rest of this paper we will investigate the case 
$(\epsilon,\lambda)=(1/g,g^2)$ in order to discuss the role of the cubic
invariant $G$.

\section{Hamiltonian structure}
The Hamiltonian structure of the model as reported in 
ref. \cite{lvov1} can be observed, by change of variables, 
$v_n = (-\epsilon)^{-n/2}u_n$,

\begin{equation}
\dot{v}_n = -i\frac{\delta [\lambda G/(-\epsilon)]}{\delta v_{-n}},
\end{equation}
where $G(v_{m},v_{-m})$ is defined as a Hamiltonian with
a density $G_n= (\lambda \sqrt{-\epsilon})^nv_{n-1}v_nv_{-(n+1)}$, which
is local in wavenumber space. This relation more a curiosity than of
practical importance. For shell models it would 
be more natural to have a Hamiltonian associated with the energy
as in most conservative dynamical systems. The attempt to construct
this has, however, until now not been fruitful \cite{Kadanoff}.

\section{The non-linear fluxes}
The non-linear transfers of the invariants are defined
as the currents, $\tilde{\Pi}^E_n = (d/dt)\sum_{m \le n}E_m$,
($\nu=f=0$), and correspondingly for $H$. They are,

\begin{eqnarray}
\tilde{\Pi}^E_n&=&k_n (D_{n+1}+(1-\epsilon)D_n/\lambda)
 \label{pie} \nonumber\\
\tilde{\Pi}^H_n&=&k_n (\epsilon-1)^{-n}(D_{n+1}-D_n/\lambda),
 \label{pih} \nonumber\\
\end{eqnarray}
where $D_n = 2 Im(u_{-(n-1)}u_{-n}u_{n+1})$.
For $\epsilon=1/g$ the flux of $G$ is,

\begin{equation}
\tilde{\Pi}^G_n=(-\lambda^2/\epsilon)^n (\lambda D^{(1)}_{n+1}
-\epsilon D^{(1)}_n/\lambda-\epsilon D^{(2)}_n + D^{(3)}_n)
 \label{pig} \nonumber\\
\end{equation}
where

\begin{eqnarray}
D^{(1)}_n&=&2 Im(u_{n+2}u_{-(n+1)}u_{-(n-1)}u_{-(n-2)}) \nonumber \\
D^{(2)}_n&=&2 Im(u_{n+2}u^2_{-n}u_{-(n-1)}) \nonumber \\
D^{(3)}_n&=&2 Im(u_{n+2}u^2_{-(n+1)}u_{n-1}) \\
\end{eqnarray}

\section{Structure functions}
Scaling exponents are obtained from expressing the transfers in
terms of structure functions. As found in \cite{lvov} the equation
(\ref{du}) is invariant under the rotation, $u_n \rightarrow exp(i\theta_n)u_n$
where the phases are a Fibonacci sequence, $\theta_{n-2}+\theta_{n-1}=\theta_n$. This symmetry is a trivial consequence of the construction
of the model. The implication of the symmetry on the structure functions
is that $\langle u_{j_1}...u_{j_p} \rangle =  
exp[i(\theta_{j_1}+...+\theta_{j_p} )]\langle u_{j_1}...u_{j_p} \rangle$.
Thus only structure functions
with $\theta_{j_1}+...+\theta_{j_p}=0$ can be non-zero. Since the phases fulfill the same relations as the associated momenta we can conclude that only
structure functions where the associated momenta sum to zero are non-zero.
The corresponding symmetry in the GOY model is $\theta_{n-2}+\theta_{n-1}+\theta_{n}=0$
leading to the artificial slowly decaying modulus(3) correlation among distant shells.
This is argued in \cite{lvov} to make this model superior to the
GOY model.

The non-vanishing structure functions can easily be listed, thus
we have the following 2nd, 3rd and 4th order structure functions,

\begin{eqnarray}
S_2(n)&=&\langle u_nu_{-n}\rangle = \langle E_n\rangle\\
S_3(n)&=&2Im\langle u_{n-1}u_{n}u_{-(n+1)}\rangle =\langle D_n\rangle\label{s3} \\
S_4^{(0)}(n,m)&=&\langle |u_n|^2|u_m|^2\rangle= \langle E_nE_m\rangle \\
S_4^{(1)}(n)&=&2Im\langle u_{n-2}u_{n-1}u_{n+1}u_{-(n+2)}\rangle=-\langle D^{(1)}_n\rangle \\
S_4^{(2)}(n)&=&2Im\langle u_{n-2}u_{-n}^2u_{n+1}\rangle=-\langle D^{(2)}_n\rangle \\
S_4^{(3)}(n)&=&2Im\langle u_{n-2}u_{n-1}^2u_{-(n+1)}\rangle=-\langle D^{(3)}_n\rangle \\
\end{eqnarray}

\section{Exact scaling relations}
The exact scaling relations corresponding to the four-fifth law simply 
states that the 
mean non-linear transfers
$\Pi^{E,H,G}_n =\langle \tilde{\Pi}^{E,H,G}_n\rangle$ are independent of
wave number within the respective inertial ranges in the high Reynolds
number limit, $\Pi^{E}_n=\overline{\epsilon}$, $\Pi^{H}_n=\overline{\delta}$ and
$\Pi^{G}_n=\overline{\eta}$, where $\overline{\epsilon},\overline{\delta},\overline{\eta}$ are
the mean dissipations of $E, H, G$ respectively.
This is usually expressed in terms of
structure functions. 
From (\ref{pie}) and (\ref{s3}) we readily obtain,

\begin{eqnarray}
k_n S_3(n+1)+ k_{n-2} S_3(n) =\overline{\epsilon} \label{s3eq}\\
k_nS_3(n+1)-k_{n-1}S_3(n)=(-1)^nk_n^{-1}\overline{\delta}.
\end{eqnarray}
in the inertial range, with the solution,

\begin{equation}
S_3(n)=\frac{\lambda^2}{k_n(1+\lambda)}(\overline{\epsilon}+(-1)^nk_n^{-1}\overline{\delta}).
\label{45}
\end{equation}
The second term on the right hand side is an oscillatory term 
which is subleading in the scaling with $k_n$ in comparison to the first term. This term
disappears 
in the case of a 'helicity free' forcing where $\overline{\delta}=0$
and we recover the shell
model correspondence to the 4/5th law expressed in terms
of the third order structure function, $S_3(n)=\overline{\epsilon}/k_n$. 

The equation similar to (\ref{s3eq}) for the transfer of the
third order quantity, $G$, reads,

\begin{eqnarray}
k_{n}^2(-\epsilon)^n[\lambda S_4^{(1)}(n+1) +\epsilon S_4^{(1)}(n)
/\lambda -\epsilon S_4^{(2)}(n)+S_4^{(3)}(n)]=\nonumber \\
k_{n}^{2+\alpha}F_4(n) =\overline{\eta}.
\label{s4eq}
\end{eqnarray}
where $\overline{\eta}$ is the mean dissipation of $G$, $F_4(n)$
denotes the square bracket
on the left hand side, and $\alpha=\log(-\epsilon)/\log(\lambda)=
(i\pi+\log(\epsilon))/\log(\lambda)$.
This was argued by L'vov et al. \cite{lvov1} to establish a
non-trivial calculation of a (sub-leading) scaling exponent, which in this
notation reads, $F_4(n)\sim k_n^{-2-\alpha}=k_n^{\tilde{\zeta}(4)} \Rightarrow 
\tilde{\zeta}(4)=2+(i\pi+\log(\epsilon))/\log(\lambda)$.
The imaginary part of the scaling exponent comes from the $(-1)^n$
factor which can be trivially eliminated by reformulating the model
in terms of new variables, $v_{2n} \equiv u_{2n}$ and $v_{2n+1}\equiv -u^*_{2n+1}$,
as was done by L'vov et al. \cite{lvov1}. 

\section{Inertial ranges}
The validity of the exact scaling relations depends on the existence of
inertial ranges separating the sources and sinks for the inviscid invariants
solely associated with the non-linear fluxes of the
invariants. The inertial range for the energy flux is determined by the
Reynolds number $Re$, which we here just associate with the inverse of the
viscosity $\nu$, assuming velocity at the integral scale to be of order unity.
The Kolmogorov scale is in complete analogy to the K41 theory determined
by balancing dissipation and non-linear flux, so we have 
$K_E \sim ({\overline{\varepsilon}}/\nu^3)^{1/4}$ growing as $Re^{3/4}$.
A similar analysis can be done for the analogous Kolmogorov scale for
dissipation of helicity \cite{D&G}. Balancing dissipation and helicity input
using $u_n \sim (\overline{\varepsilon}/k_n)^{1/3}$ gives, $K_H\sim
[{\overline{\delta}}^3/(\nu^3{\overline{\varepsilon}}^2)]^{1/7}$. Thus we
get $K_H/K_E \sim \nu^{-3/7+3/4}=\nu^{9/28} \rightarrow 0$ for $\nu\rightarrow 0$. 
This means that for high Reynolds number flow the small scales will always
exhibit non-helical flow. In the shell model the helicity changes sign due to the $(-1)^n$ factor.
The dissipation of positive helicity at even numbered shells and negative
helicity at odd numbered shells will grow with wavenumber as $D^H_n \sim (-1)^n k_n^3 |u_n|^2 \sim
(-1)^n k_n^{7/3}$, consequently
the shell model will show strong odd-even oscillations 
of $\Pi^H_n$ from balancing the positive and negative dissipations, the scaling will
be determined by the dissipation $|\Pi^H_n|\sim k_n^{7/3}$ for $k_n>K_H$. 
The situation for the cubic invariant $G$ is different. We can again define a
Kolmogorov scale for dissipation of $G$ by equating the 
the dissipation and the input of $G$, 

\begin{equation}
\nu k_n^2 G_n \sim \nu (-1)^n k_n^{7/2} {\cal R}[u_{n-1}u_nu^*_{-(n+1)}] \sim {\overline{\eta}}
\end{equation}
using $u_n \sim (\overline{\varepsilon}/k_n)^{1/3}$ again gives,

\begin{equation}
K_G \sim [\overline{\eta}/(\overline{\varepsilon}\nu)]^{2/5}.
\end{equation}
The ratio of dissipation scales is then $K_G/K_E \sim \nu^{-2/5+3/4} =\nu^{7/20}$, so
as for the case of helicity the small scales will have no net non-linear flux of $G$.
There is, however, a crucial difference between the dissipation of $H$ and of $G$. The
dissipation of helicity is forced to be of alternating signs whereas the dissipation 
of $G$ can be of either sign at any shell. Thus one should expect the mean dissipation
of $G_n$ to vanish for $k_n > K_G$, and the dissipation would not determine the 
scaling of $\Pi^G_n$. So how would the scaling of $\Pi^G_n$ be then? The
non-linear flux $\Pi^G_n$ is constituted of terms of the form $k_n^{5/2}S_4(n)\sim k_n^{5/2}u_n^4
\sim k_n^{7/6}$. This leading scaling is killed by detailed cancellations between terms in an
inertial range.
One would then
expect $K_G$ to represent a de-correlation scale where the individual terms become
independent and the scaling of $|\Pi^G_n|$ becomes approximately that of the individual
terms, which is $k_n^{7/6}$. This conjecture is tested in a numerical simulation.
 
\section{numerical simulation}
A simulation using a simple fourth order Runge-Kutta scheme 
has been performed with the parameter values, $(\epsilon, \lambda, \nu, N, f)
= (1/g,g^2,10^{-9},20,(1+i)\delta_{1,n})$. The anomalous scaling exponents are defined from
$\langle (\tilde{\Pi}^E_n)^{p/3}\rangle \sim k_n^{\zeta(p)}$. The results for this
model are show in figure 2 (diamonds) and coincides within the numerical 
accuracy of the simulation with the exponents 
found by L'Vov et al. for the Sabra model with $(\epsilon,
\lambda)=(1/2,2)$ (triangles). The non-linear fluxes for $E, H, G$ are shown in figures 3, 4 and 5. 
The fluxes for $H$ and $G$ fluctuates between positive and negative values, so that only
the absolute values are shown. The scaling indicated by the straight lines confirms
the conjectures made in the previous section. 
However, even though the simulation is long enough and numerically accurate
enough to determine the anomalous scaling exponents reliably, the non-linear
flux $\Pi^G_n$ can only be determined using extreme numerical precision. If we conjecture
that there is a constant flux of $G$ of order unity $(\overline{\eta} \sim 1)$
through the inertial range we must
calculate $\Pi^G_n$ as differences of fourth order correlators of order $k_n^{7/6}$
which is about $10^6$ at the end of the inertial range to obtain a constant of order
unity. So in fact the graph in figure 5 does not reliably represent $\Pi^G_n$, 
it is merely numerical noise. To see this consider the average 
$|\langle \Pi^G_n \rangle (t)|=|\int^t_0 \Pi^G_n(\tau)d\tau|$ as a function of $t$,
see figure 6. The line is the curve $\sigma /\sqrt{t}$ which is expected for an independent 
random process. The standard deviation $\sigma$ of the process is of the order $k_n^{7/6}$.
The main justification for studying shell models is the possibility of
accurate numerical calculations of correlators and scaling exponents for high
Reynolds number flow.
Here we see that even the shell model can been pushed to the limit where the determination
of correlators by numerical simulation
is impractical. 

\section{summary}
To summarize, it has been argued why this model is a 
natural choice for a shell model of turbulence. For the choice of parameters 
which conserves both energy, dimensionally correct helicity and the 
third order quantity $G$ the anomalous scaling exponents are the same as for 
the Sabra model which conserves energy and dimensionally correct helicity but
not $G$. 
So even
though the model has a Hamiltonian structure, the 
non-positive Hamiltonian, $G$, seems not relevant for
determining the scaling properties of the model.
It has been argued that there will not be an inertial range scaling regime 
for the fourth order correlator associated with the flux of $G$. However, 
due to the extreme numerical accuracy required to determine this correlator 
the numerical simulations presented here are not conclusive in the determination
of this correlator. 

\section{Acknowledgement}
I would like to thank M. H. Jensen and P. Giuliani
for valuable discussions. The work was funded
by the Carlsberg foundation.


\newpage
\begin{center}
FIGURE CAPTIONS
\end{center}
\newcounter{fig}
\begin{list}{Fig. \arabic{fig}}
{\usecounter{fig}\setlength{\labelwidth}{2cm}\setlength{\labelsep}{3mm}}
\item The parameter space $(\epsilon, \lambda)$ for the shell model. Energy is by
construction always conserved. The hyperbola 
corresponds to conservation of dimensionally correct helicity $H=\sum_n (-1)^n k_n |u_n|^2$
the horizontal line is the golden ratio shell spacing where the 'triangles' close, 
$k_{n-2}+k_{n-1}=k_n$. The vertical line represents parameters for which $G=\sum_n
(-\lambda/\epsilon)^n {\cal R}[u_{n-1}u_nu_{-(n+1)}]$ is an inviscid invariant.
At the cross 'helicity' has the form $H=\sum_n (-1)^n k_n^{1/2} |u_n|^2$.

\item The anomalous scaling exponents $\zeta(p)$ defined from $(\Pi^E_n/k_n)^{p/3}\sim k_n^{-\zeta(p)}$  for $(\epsilon, \lambda)=(1/g,g^2)$ (diamonds) coincides within the numerical uncertainty
with the values found by L'Vov at al. for $(\epsilon, \lambda)=(1/2, 2)$ (triangles).

\item The non-linear flux of energy $\Pi^E_n$ as a function of $k_n$.
\item The non-linear flux of helicity $|\Pi^H_n|$ as a function of $k_n$. An inertial range
with coexisting cascades of both energy and helicity is seen for the first few
shells, after which the dissipation of helicity dominates the spectrum. The
sign of the flux alternates for even and odd shells corresponding to dissipation
of positive and negative helicity. The straight line indicates the scaling exponent
$7/3$ as is expected from the dissipation.

\item The non-linear flux of the third order quantity $\Pi^E_n$ as a function of $k_n$.
The numerics has not converged and the scaling exponents $7/6$ indicated by the line
comes from the standard deviation of the numerical value of the difference between
correlators of the order $k_n^{7/6}$.

\item The average $|\langle \Pi^G_n \rangle (t)|=|\int^t_0 \Pi^G_n(\tau)d\tau|$ as a function of $t$.
The straight line has the slope $-1/2$ as for an independent random variable.
This shows that the quantity shown in figure 5 is dominated by noise. Note that the 
sampling time is long enough to determine the anomalous scaling exponents.
\end{list}

\newpage
\begin{figure}[htb]
\epsfxsize=10cm
\epsffile{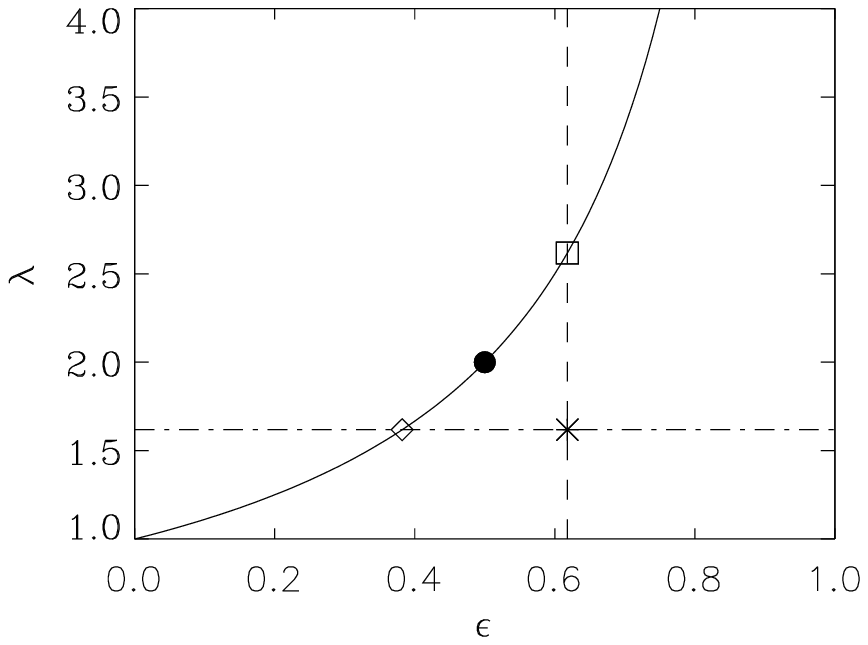}
\caption[]{
}
\end{figure}

\begin{figure}[htb]
\epsfxsize=10cm
\epsffile{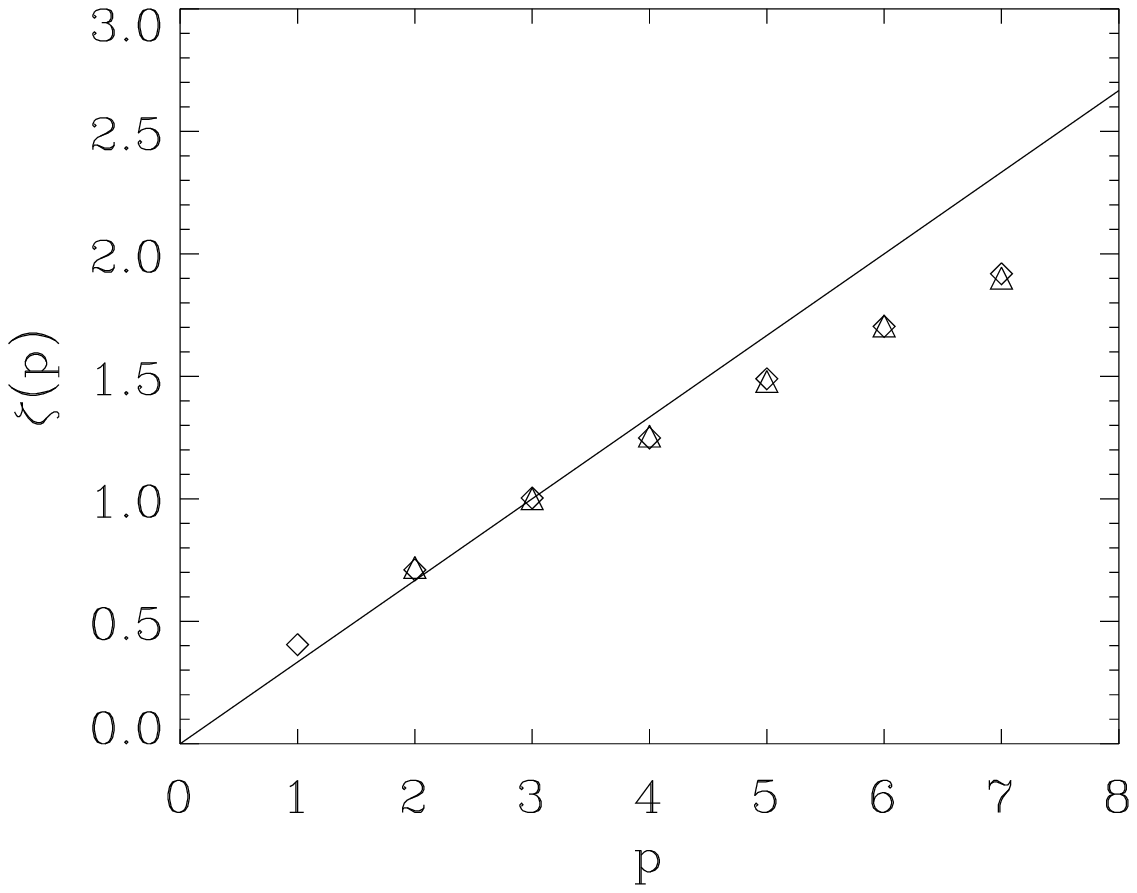}
\caption[]{
 }
\end{figure}

\begin{figure}[htb]
\epsfxsize=10cm
\epsffile{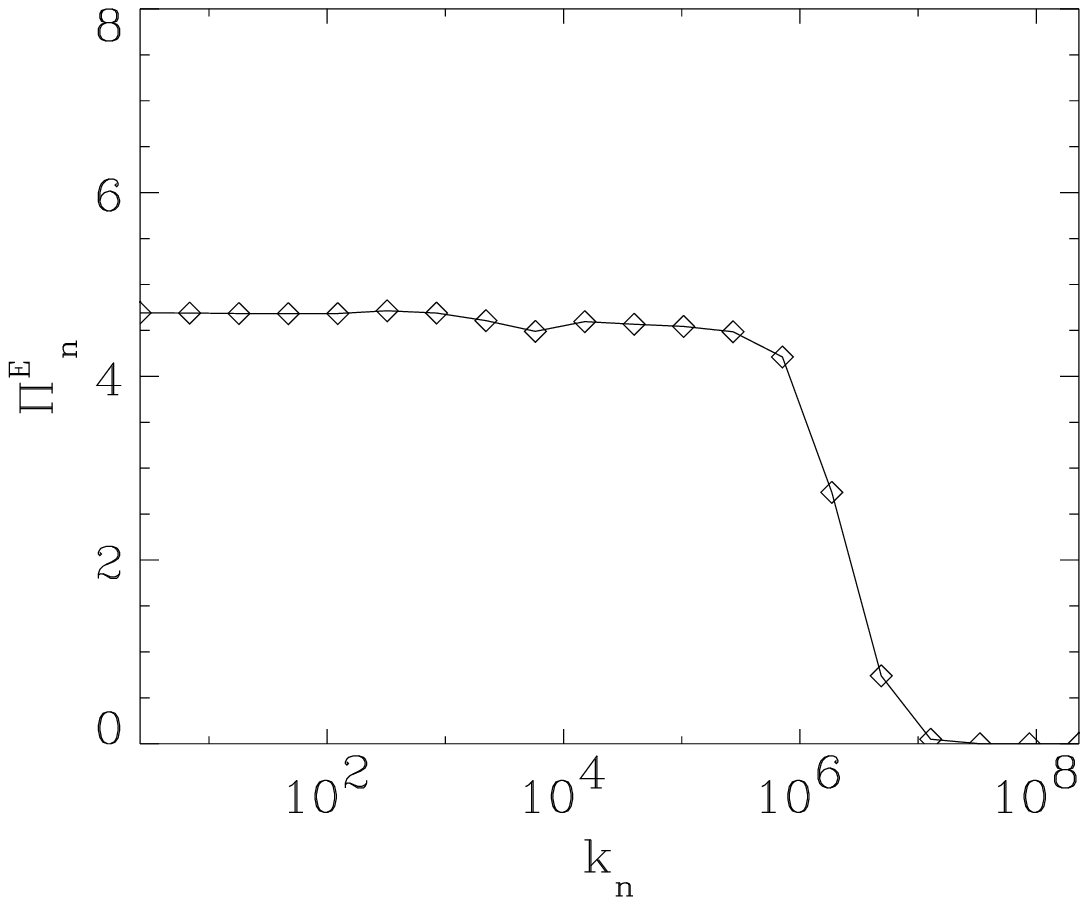}
\caption[]{
 }
\end{figure}

\begin{figure}[htb]
\epsfxsize=10cm
\epsffile{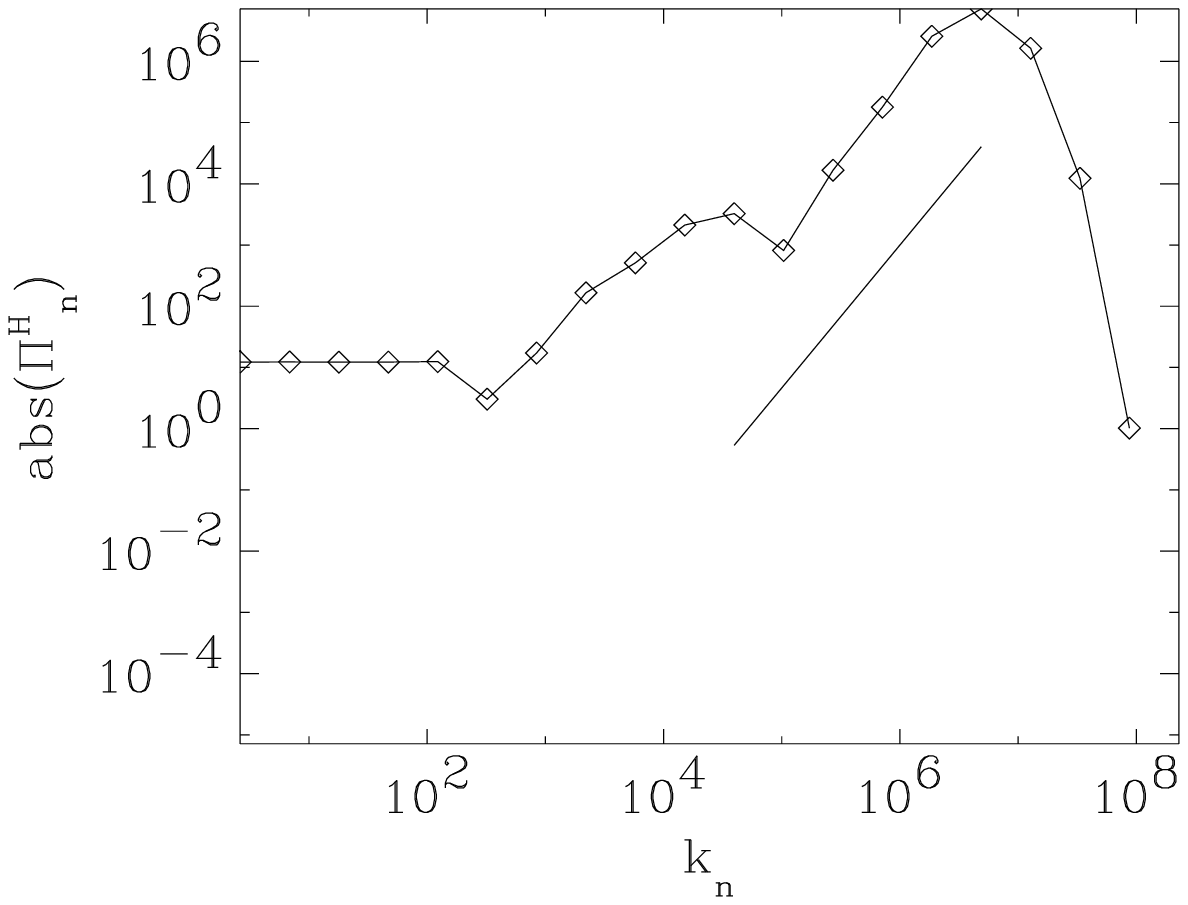}
\caption[]{
 }
\end{figure}

\begin{figure}[htb]
\epsfxsize=10cm
\epsffile{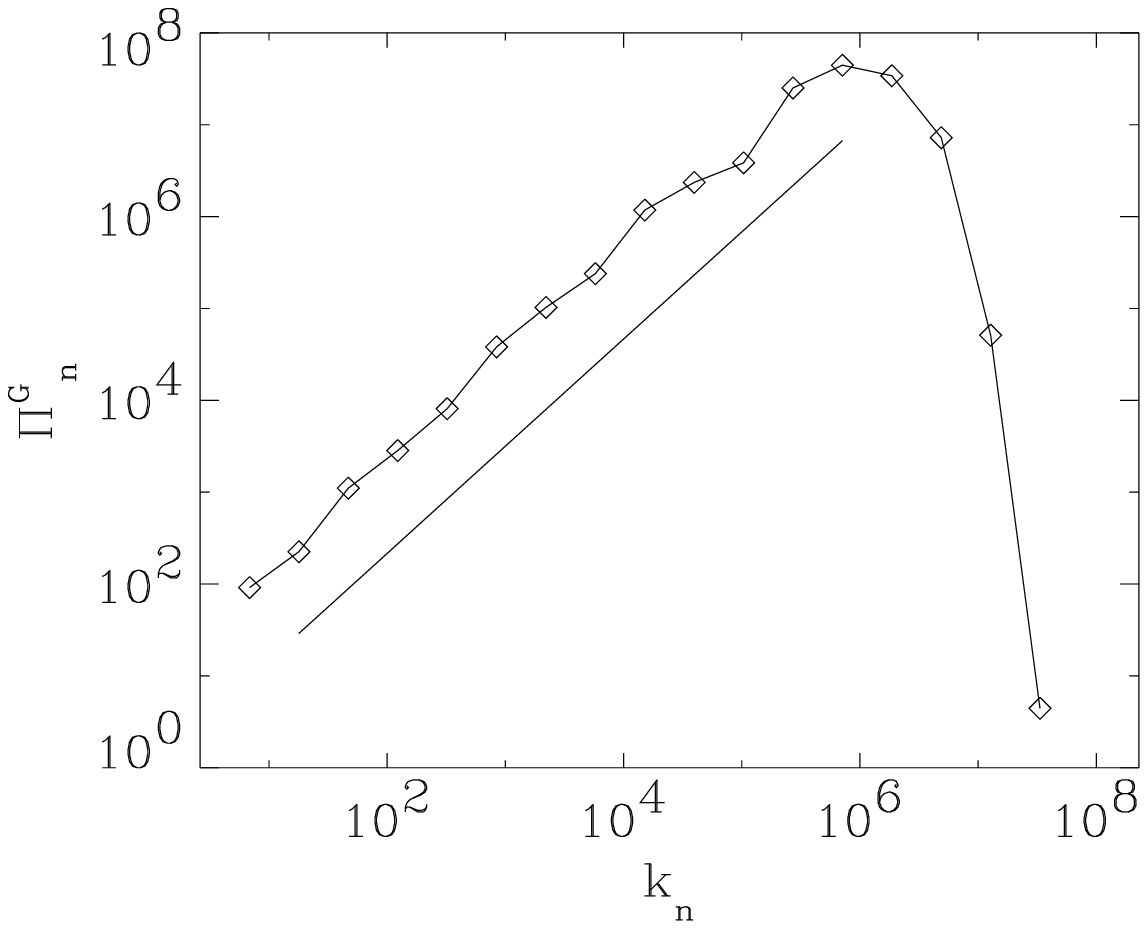}
\caption[]{
 }
\end{figure}

\begin{figure}[htb]
\epsfxsize=10cm
\epsffile{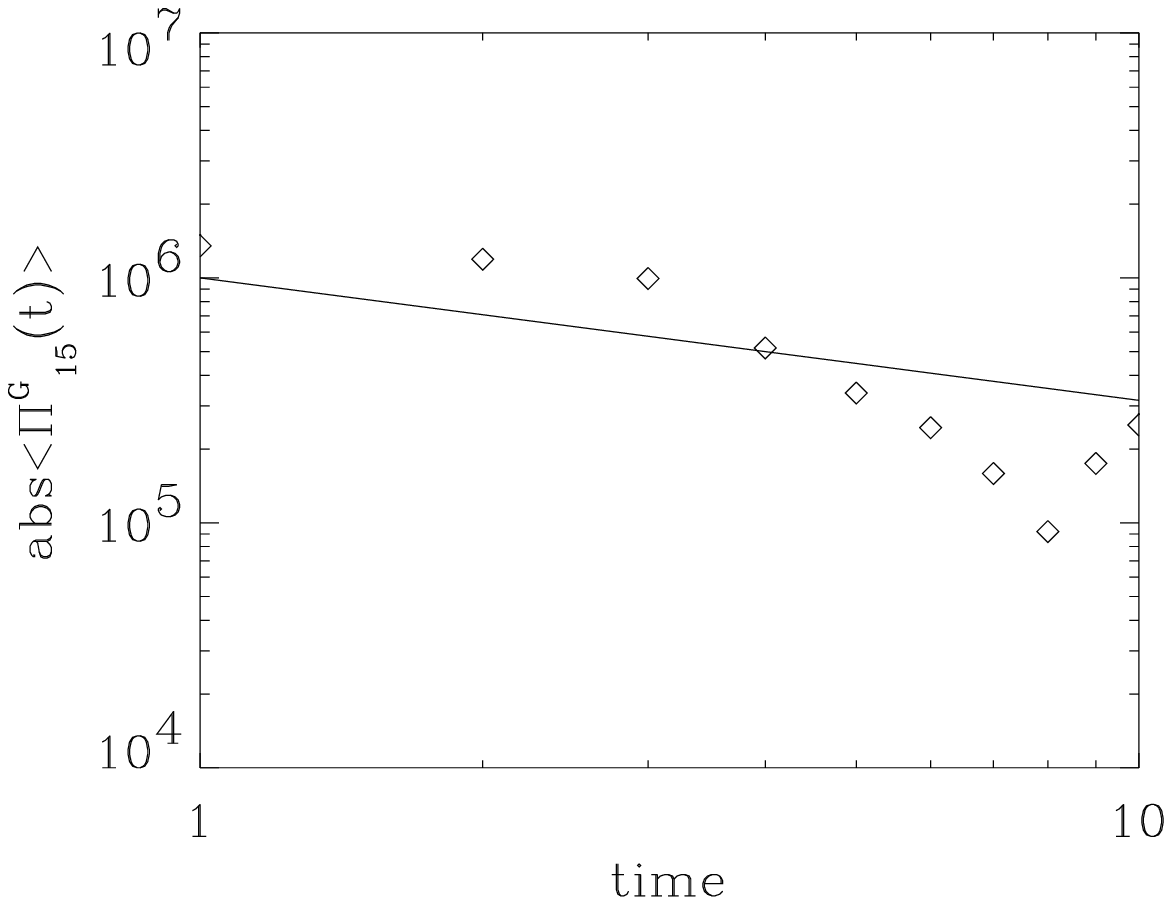}
\caption[]{
 }
\end{figure}

\end{document}